\documentclass{eptcs}

\usepackage{breakurl}
\usepackage{abbreviations}

\usepackage[normalem]{ulem}

\usepackage{ifthen,stmaryrd}
\usepackage{amsmath, stmaryrd, amssymb,mathtools}
\usepackage{graphicx,xspace}
\usepackage{url}
\usepackage{cancel}
\usepackage{theorem}
\usepackage[usenames,dvipsnames]{color}
\usepackage{subfigure}
\usepackage{tikz}
\usetikzlibrary{arrows,automata}
\usepackage{proof}

\newtheorem{theorem}{Theorem}
\newtheorem{definition}{Definition}
\newtheorem{example}{Example}

\newtheorem{algorithm}{Algorithm}

\title{Decomposability in Input Output Conformance Testing}

\author{Neda Noroozi
\institute{Eindhoven University of Technology\\ Eindhoven, The Netherlands}
\email{n.noroozi@tue.nl}
\and
Mohammad Reza Mousavi
\institute{Eindhoven University of Technology\\ Eindhoven, The Netherlands}
\institute{Center for Research on Embedded Systems (CERES)\\ Halmstad University, Sweden}
\email{m.r.mousavi@tue.nl}
\and
Tim A.C. Willemse
\institute{Eindhoven University of Technology\\ Eindhoven, The Netherlands}
\email{t.a.c.willemse@tue.nl}
}

\def\titlerunning{Decomposability in IOCO}
\def\authorrunning{Noroozi, Mousavi \&  Willemse}

\begin{document}
\maketitle

\begin{abstract}
We study the problem of deriving a specification for a third-party component,
based on the specification of the system and the environment in which the component is supposed to reside.
Particularly, we are interested in using component specifications for conformance testing of black-box components, using the theory of input-output conformance (ioco) testing.
We propose and prove sufficient criteria for decompositionality, i.e., that components conforming to the derived specification will always compose to produce a correct system with respect to the system specification.
We also study the criteria for strong decomposability, by which we can ensure that only those components conforming to the derived specification can lead to a correct system.
\end{abstract}

\section{Introduction}\label{sec:intro}

Enabling reuse and managing complexity are among the major benefits of using
compositional approaches in software and systems engineering.  This idea
has been extensively adopted in several different subareas of software
engineering, such as product-line software engineering.  One of the
cornerstones of the product-line approach is to reuse a common platform
to build different products. This common platform should ideally comprise
different types of artifacts, including test-cases, that can be re-used
for various products of a given line.  In this paper, we propose an
approach to conformance testing, which allows to use a high-level
specification and derive specifications for to-be-developed components
(or sub-systems) given the platform on which they are to be deployed.
We call this approach \emph{decompositional} testing and refer to the
process of deriving  specifications as \emph{quotienting} (inspired by
its counterpart in the domain of formal verification).

We develop our approach within the context of input-output conformance
testing (\ioco) \cite{Tretmans08}, a model-based testing theory
using formal models based on input-output labeled transition systems
(\iolts{}s).  An implementation $i$ is said to conform to a specification
$s$, denoted by $i ~\ioco~ s$, when after each trace in the specification,
the outputs of the implementation are among those prescribed by the
specifications.

\begin{figure}[ht]
\label{fig:structure}
  \centering
\begin{tikzpicture}[scale=0.7]
\tikzstyle help lines=[color=gray!20,very thin]
    \draw (1.5,0)rectangle (10.5,2);
    \draw [thick] (2,0.25) rectangle node {platform $\bar{e}$}(5.5,1.75);
    \draw [thick] (6.5,0.25) rectangle node  {component $\bar{c}$}(10,1.75);
    \draw [arrows={triangle 45-}](0,0.5) -- node[above]  {$U'_e$}(2,.5);
    \draw [arrows={-triangle 45}](0,1.37) -- node[above]  {$I'_e$}(2,1.37);
    \draw [arrows={triangle 45-}](5.5,.5) -- node[above]  {$U_v$}(6.5,.5);
    \draw [arrows={-triangle 45}](5.5,1.37) -- node[above]  {$I_v$}(6.5,1.37);
    \draw [arrows={-triangle 45}](10,.5) -- node[above]  {$U'_c$}(12,.5);
    \draw [arrows={triangle 45-}](10,1.37) -- node[above]  {$I'_c$}(12,1.37);
\end{tikzpicture}
\caption{Strucure of a system composed of platform $\bar{e}$ and
component $\bar{c}$ whose behavior is defined by a given specification
$\bar{s}$. The language of platform $\bar{e}$ comprises $(I'_e\cup U_v)\cup (U'_e\cup I_v)$. 
Similarly, $(I'_c\cup I_v)\cup (U'_c\cup U_v)$ is the language of component $\bar{c}$. The platform $\bar{e}$ and component $\bar{c}$ interface
via $I_v$ and $U_v$ which are hidden from the viewpoint of an external observer.}
\end{figure}
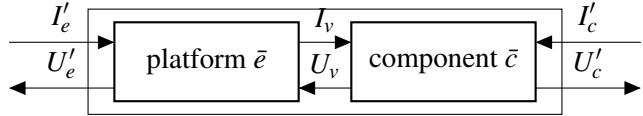

For a given platform (environment) $\bar{e}$, whose behavior
is given as an \iolts,  a quotient of a specification $\bar{s}$ by the
platform $\bar{e}$, denoted by $\quotient{\bar{s}}{\bar{e}}$, is the specification
that describes the system after filtering out the effect
of $\bar{e}$.  The structure of a system consisting of $\bar{e}$ and
unknown component $\bar{c}$ is represented in Figure \ref{fig:structure},
whose behavior is described by a given specification $\bar{s}$.  We would like to construct
$\quotient{\bar{s}}{\bar{e}}$ such that it captures the behavior of any
component $\bar{c}$ which, when deployed on $\bar{e}$ (put in parallel and
possibly synchronize with $\bar{e}$) conforms to $\bar{s}$. Put formally,
$\quotient{\bar{s}}{\bar{e}}$ is the specification which satisfies the
following bi-implication:

\[
\forall \bar{c},\bar{e}.~~ \bar{c} ~\ioco~ \quotient{\bar{s}}{\bar{e}} ~~~~\Leftrightarrow~~~~ \pcomposition{\bar{c}}{\bar{e}} ~\ioco~ \bar{s}
\]
The criteria for the implication from left to right, which is essential
for our approach, are called \emph{decomposability}.  The criteria for
the implication from right to left guarantee that quotienting produces
the precise specification for the component and is called \emph{strong
decomposability}.  We study both criteria in the remainder of this paper.
Moreover, we show that strong decomposability can be combined with
\emph{on-the-fly} testing, thereby avoiding constructing the
witness to the decomposability explicitly upfront.

\paragraph{Related Work.}
The study of compositional and modular verification  for various temporal and modal logics has attracted considerable attention and several compositional verification techniques have been proposed for
such logics; see, e.g., \cite{Berezin98,Kupferman98,Pasareanu99,Giannakopoulou05}.  Decompositional reasoning aims at automatically decomposing the global property to be model checked into local
properties of (possibly unknown) components, a technique that is often called quotienting.
The notion of quotient introduced in the present paper is inspired by its corresponding 
notion in the area of (de)compositional model-checking,
and is substantially adapted to the setting for input-output conformance testing, e.g., 
by catering for the distinction between input and output actions and 
taking care of (relative) quiescence of components. 
In the area of model-based testing, we are aware of a few studies dedicated to the issue of (de)composition \cite{BijlRT03,FrantzenT06,unknownComponent}, 
of which  we give an overview below.

In \cite{BijlRT03} the compositionality of the \ioco-based testing
theory is investigated. Assuming that implementations of components
conform to their specifications, the authors investigate whether the
composition of these implementations still conforms to the composition
of the specifications. They show that this is not necessarily the case
and they establish conditions under which \ioco is a compositional testing
relation.


In~\cite{FrantzenT06}, Frantzen and Tretmans study when successful
integration of components by composing them in certain ways can be
achieved.  Successful integration is determined by two conditions:
the integrated system correctly provides services, and interaction
with other components is proper.  For the former, a specification
of the provided services of the component is assumed.  Based on the
\ioco-relation, the authors introduce a new implementation relation
called $\textbf{eco}$, which allows for checking whether a component
conforms to its specification as well as whether it uses other components
correctly. In addition, they also propose a bottom-up strategy for
building an integrated systems.

Another problem closely related to the problem we consider in this
paper is \emph{testing in context}, also known as \emph{embedded
testing}~\cite{unknownComponent}.  In this setting,
the system under test comprises a component $\bar{c}$ which is embedded in a
context $\bar{u}$. Component $\bar{c}$ is isolated from the
environment and all its interactions proceed through
$\bar{u}$ (which is assumed to be correctly implemented). The implementation $\bar{i}$ and specification $\bar{s}$
of the system composed of $\bar{u}$ and $\bar{c}$, are assumed to be
available. The problem of testing in context then entails generating a
test suite that allows for detecting incorrect implementations $\bar{i}$
of component $\bar{c}$.

Although testing in context and decomposability share
many characteristics, there are key differences between the two.  We do
not restrict ourselves to embedded components, nor do we assume the
platforms to be fault-free.  Contrary to the testing in context approach,
decomposing a monolithic specification is the primary challenge in our
work; testing in context already assumes the specification is the result
of a composition of two specifications. Moreover, in testing in context,
the component $\bar{c}$ is tested through context $\bar{u}$ whereas
our approach allows for testing the component directly
through its deduced specification.  As a result, we do not require that
the context is always available while testing the component,
which is particularly important in case the
platform is a costly resource.

For similar reasons, asynchronous testing
\cite{SimaoP11,Neda11,WeiglhoferW09}, which can be considered
as some form of embedded testing, is different from the work we present
in this paper.

\paragraph{Structure.}
We give a cursory overview of \ioco-based formal testing in
Section~\ref{sec::pre}.  The notions of decomposability and strong
decomposability are formalized in Section~\ref{sec::decompos}.
We present sufficient conditions for determining whether a given specification
is decomposable in Section~\ref{sec::sync} and whether it is strongly
decomposable in Section~\ref{sec::strong_decompos}. We conclude in
Section \ref{sec::conc}. \textbf{\emph{Additional examples and results, together with all
proofs for the lemmata and theorems can be found in~\cite{NMW:12}.}}

\section{Preliminaries}\label{sec::pre}

Conformance testing is about checking that the observable behavior of
the system under test is included in the prescribed behavior of the
specification. In order to formally reason about conformance testing,
we need a model for reasoning about the behaviors described by a
specification, and assume that we have a formal model representing the
behaviors of our implementations, so that we can reason about their
conformance mathematically.

In this paper, we use variants of the
well-known Labeled Transition Systems as a behavioral model for both
the specification and the system under test.
The Labeled Transition System model assumes that systems can be
represented using a set of states and transitions, labeled with events
or \emph{actions}, between such states.  A tester can observe the events
leading to new states, but she cannot observe the states. We assume
the presence of a special action $\tau$, which we assume is unobservable
to the tester.
\begin{definition}[\iolts]\label{def:IOLTS}
An input-output labeled
transition system (\iolts) is a  tuple $\langle S, I, U, \rightarrow,
\bar{s} \rangle$, where $S$ is a set of states, $I$ and $U$ are disjoint
sets of observable \emph{inputs} and \emph{outputs}, respectively,
$\rightarrow \subseteq S \times (I \cup U \cup \{\tau\}) \times S$ is the
transition relation (we assume $\tau \notin I \cup U$), and $\bar{s} \in S$
is the initial state.  The class of \iolts{}s ranging over
inputs $I$ and outputs $U$ is denoted $\IOLTS{I,U}$.
\end{definition}

Throughout this section, we assume an arbitrary, fixed \iolts $\langle S,
I, U, \to, \bar{s} \rangle$, and we refer to this \iolts by referring to
its initial state $\bar{s}$.  We write $L$ for the set $I\cup U$.  Let $s,
s'\in S$ and $x \in L \cup \{\tau\}$. In line with common practice,
we write $s \trans{x} s'$ rather than $(s,x,s') \in \to$. Furthermore,
we write $s \trans{x}$ whenever $s \trans{x} s'$ for some $s' \in S$,
and $s \ntrans{x}$ when not $s \trans{x}$. A \emph{word} is a sequence
over the input and output symbols. The set of all words over $L$ is
denoted $L^*$, and $\ep$ is the empty word. For words $\sigma, \rho
\in L^*$, we denote the concatenation of $\sigma$ and $\rho$ by $\sigma
\rho$. The transition relation is generalized to a relation over words
by the following deduction rules:
\[
\begin{array}{ccccc}
\infer{s \Trans{~\epsilon~} s}{}
&
\qquad
&
\infer{s \Trans{~\sigma x~} s'}
      {s \Trans{~\sigma~} s'' \qquad s'' \trans{~x~} s' \qquad x \not=\tau}
&
\qquad
&
\infer{s \Trans{~\sigma} s'}
      {s \Trans{~\sigma~} s'' \qquad s'' \trans{~\tau~} s'}
\end{array}
\]
We adopt the notational conventions we introduced for $\rightarrow$ for
$\Longrightarrow$.
A state in the \iolts $\bar{s}$ is
said to \emph{diverge} if it is the source of an infinite sequence of
$\tau$-labeled transitions. The \iolts $\bar{s}$ is \emph{divergent} if
one of its reachable states diverges.
Throughout this paper, we confine ourselves to non-divergent \iolts{}s.
%


\begin{definition} Let $s' \in S$ and $S' \subseteq S$.
The set of \emph{traces}, \emph{enabled actions} and
\emph{weakly enabled actions} for $s$ and $S'$ are defined as follows:
\begin{itemize}

\item $\traces{s} = \{ \sigma \in L^* ~|~ s \Trans{\sigma} \}$, and
$\traces{S'} = \bigcup\limits_{s' \in S'} \traces{s'}$.

\item $\init{s} = \{ x \in L \cup \{\tau\} ~|~ s \trans{x} \}$, and
$\init{S'} = \bigcup\limits_{s' \in S'} \init{s'}$.

\item $\Sinit{s} = \{ x \in L  ~|~ s \Trans{x} \}$, and
$\Sinit{S'} = \bigcup\limits_{s' \in S'} \Sinit{s'}$.

\end{itemize}

\end{definition}

\paragraph{Quiescence and Suspension Traces.} Testers often
not only have the power to observe events produced by an
implementation, they can also observe the \emph{absence} of events,
or \emph{quiescence}~\cite{Tretmans08}. A state $s \in S$ is said
to be \emph{quiescent} if it does not produce outputs and it is
\emph{stable}. That is, it cannot, through internal computations,
evolve to a state that \emph{is} capable of producing outputs. Formally,
state $s$ is quiescent, denoted $\quiescent{s}$, whenever $\init{s}
\subseteq I$.
In order to formally reason about the observations of inputs, outputs and
quiescence, we introduce the set of \emph{suspension
traces}. To this end, we first generalize the transition relation over
words to a transition relation over suspension words.  Let
$L_\delta$ denote the set $L \cup \{\delta\}$.

\[
\begin{array}{ccccc}
\infer{s \Trans{~\sigma~}_\delta s'}{s \Trans{~\sigma~} s'}
&
\qquad
&
\infer{s \Trans{~\delta}_\delta s}
      {\quiescent{s}}
&
\qquad
&
\infer{s \Trans{~\sigma \rho~}_\delta s'}
      {s \Trans{~\sigma~}_\delta s'' \qquad s'' \Trans{~\rho~}_\delta s'}

\end{array}
\]

The following definition formalizes the set of suspension traces.
\begin{definition} Let $s \in S$ and $S' \subseteq S$. The set
of \emph{suspension traces} for $s$, denoted
$\Straces{s}$ is defined as the set $\{\sigma \in L_\delta^* ~|~
s \Trans{\sigma}_\delta \}$; we set $\Straces{S'} = \bigcup\limits_{s' \in S'}
\Straces{s'}$.

\end{definition}

\paragraph{Input-Output Conformance Testing with Quiescence.}

Tretmans' \ioco testing theory~\cite{Tretmans08} formalizes black
box conformance of implementations. It assumes that the behavior of
implementations can always be described adequately using a class of
\iolts{}s, called \emph{input output transition systems}; this assumption
is the so-called \emph{testing hypothesis}.  Input output transition
systems are essentially plain \iolts{}s with the additional assumption
that inputs can always be accepted.

\begin{definition}[\iots]\label{def:IOTS} Let $\langle S, I,
U, \rightarrow, \bar{s} \rangle$ be an \iolts. A state $s \in S$ is
\emph{input-enabled} iff $I \subseteq \Sinit{s}$; the \iolts $\bar{s}$
is an \emph{input output transition system} (\iots) iff every state $s
\in S$ is input-enabled. The class of input output transition systems
ranging over inputs $I$ and outputs $U$ is denoted $\IOTS{I, U}$.

\end{definition}
While the \ioco testing theory assumes input-enabled implementations,
it does not impose this requirement on specifications. This facilitates
testing using partial specifications, \emph{i.e.}, specifications that
are under-specified. We first introduce the main concepts that are used
to define the family of conformance relations of the \ioco testing theory.
\begin{definition} Let $\langle S, I, U, \rightarrow, \bar{s} \rangle$ be an
\iolts. Let $s \in S$, $S' \subseteq S$ and let
$\sigma \in L_\delta^*$.

\begin{itemize}

\item $s \after \sigma = \{ s' \in S ~|~ s \Trans{\sigma}_\delta s' \}$,
and $S' \after \sigma = \bigcup\limits_{s' \in S'} s' \after \sigma$.

\item $\out{s} = \{ x \in L_\delta \setminus I ~|~ s \Trans{x}_\delta \}$,
and $\out{S'} = \bigcup\limits_{s' \in S'} \out{s'}$.

\end{itemize}

\end{definition}
The family of conformance relations for \ioco are then defined as follows,
see also~\cite{Tretmans08}.

\begin{definition}[\ioco]\label{def:IOCO}
Let $\langle R, I, U, \rightarrow, \bar{r} \rangle$ be an \iots representing
a realization of a system, and
let \iolts $\langle S, I,U, \rightarrow, \bar{s} \rangle$
be a specification. Let $F \subseteq L_\delta^*$. We say
that $\bar{r}$ is \emph{input output conform} with specification $\bar{s}$, denoted
$\bar{r} \IOCO{F} \bar{s}$, iff
\[
\forall \sigma \in F:~ \out{\bar{r} \after \sigma}
\subseteq \out{\bar{s} \after \sigma}
\]

\end{definition}
The $\IOCO{F}$ conformance relation can be specialized by choosing an
appropriate set $F$. For instance, in a setting with $F = \Straces{s}$, we obtain the \ioco relation originally
defined by Tretmans in~\cite{Tretmans96}.  The latter conformance
relation is known to admit a sound and complete test case generation
algorithm, see, \emph{e.g.},~\cite{Tretmans96,Tretmans08}. Soundness
means, intuitively, that the algorithm will never generate a test case
that, when executed on an implementation, leads to a \emph{fail} verdict
if the test runs are in accordance with the specification. Completeness
is more esoteric: if the implementation has a behavior that is not in
line with the specification, then there is a test case that, in theory,
has the capacity to detect that non-conformance.
\paragraph{Suspension automata.}
\label{sec::esa}

The original test case generation algorithm by Tretmans for the \ioco
relation relied on an automaton derived from an \iolts specification.
This automaton, called a \emph{suspension automaton}, shares many
of the characteristics of an \iolts, except that the observations of
quiescence are encoded explicitly as outputs: $\delta$ is treated as
an ordinary action label which can appear on a transition. In addition,
Tretmans assumes these suspension automata to be \emph{deterministic}:
any word that could be produced by an automaton leads to exactly one
state in the automaton.

\begin{definition}[Suspension automaton]\label{def:SA}
A suspension automaton(SA) is a deterministic $\iolts$ $\langle S, I,
U \cup \{\delta\}, \to, \bar{s} \rangle$;
that is, for all $s \in S$ and all $\sigma \in L^*$, we have
$|s \after \sigma | \le 1$.
\end{definition}
Note that determinism implies the absence of $\tau$ transitions.
In~\cite{Tretmans96}, a transformation from ordinary \iolts{}s to
suspension automata is presented; the transformation ensures that
trace-based testing using the resulting suspension automaton is
exactly as powerful as \ioco-based testing using the original \iolts.

The transformation is essentially based on the subset construction
for determinizing automata.
Given an \iolts,  the transformation $\Delta$ defined below converts any \iolts into
an SA.
\begin{definition}\label{def:trans_SA_Ex}\label{def:Delta_SA}
Let $\tuple{S, I, U, \rightarrow, \bar{s}} \in \IOLTS{I, U}$. The
SA $\Delta(\bar{s}) = \tuple{Q, I, U\cup\{\delta\}, \to, \bar{q}}$
is defined as:
\begin{itemize}
\item $Q = \powerSet{S} \setminus \{ \emptyset \}$.
\item $\bar{q} = \bar{s} \after \ep$.
\item $\to \subseteq Q \times L_\delta \times Q$ is the least
relation satisfying:
\end{itemize}
\[
\begin{array}{ccc}
\infer{q\trans{x} \{s'\in S \mid \exists s\in q\st s\Trans{x} s' \}}%
      {x\in L \qquad q\in Q}
&
\qquad
&
\infer{q\trans{\delta} \{s\in q\mid \quiescent{s} \}}
      {q \in Q}
\end{array}
\]

\end{definition}

\begin{example}
Consider the \iolts $\bar{s}$ depicted in Figure~\ref{fig:FIG} on
page~\pageref{fig:FIG}. The \iolts $\bar{s}$ is a specification of
a malfunctioning vending machine which sells tea for one euro coin
(c).  After receiving money, it either delivers tea
(t), refunds the money (r) or does nothing.
Its suspension automaton $\Delta(\bar{s})$, with initial state
$\bar{q}$, is depicted next to it.
Note that the suspension traces of $\bar{s}$ and the traces of suspension
automaton $\Delta(\bar{s})$ are identical.

\end{example}
In general, a suspension automaton may not represent an actual \iolts;
for instance, in an arbitrary suspension automaton, it is allowed to
observe quiescence, followed by a proper output. This cannot happen
in an \iolts. In~\cite{Willemse}, the set of suspension automata
is characterized for which a transformation to an \iolts \emph{is}
possible. Such suspension automata are called \emph{valid}. 
%
%
%
%
%
%
Proposition~1 of~\cite{Willemse} states that for any
\iolts $\bar{s}$, the suspension automaton $\Delta(\bar{s})$ is valid.
Conversely, Theorem~2 of~\cite{Willemse} states that any valid suspension
automaton has the same testing power (with respect to \ioco) as
\emph{some} \iolts. This essentially means that the class of valid
suspension automata can be used safely for
testing purposes.

\paragraph{Parallel Composition.}

A software or hardware system is usually composed of subunits and
modules that work in an orchestrated fashion to achieve the desired
overall behavior of the software or hardware system. In our setting,
we can formalize such compositions using a special operator
$\pcomposition{\_}{\_}$ on \iolts{}s: two \iolts{}s can interact by
connecting the outputs sent by one \iolts to the inputs of the other
\iolts. We assume that such inputs and outputs are taken from a shared
alphabet of actions. For the non-common actions the behavior of both
\iolts{}s is interleaved.

\begin{definition}[parallel composition]\label{def:pcomposition} Let
$\langle S_1, I_1, U_1, \rightarrow_1, \bar{s}_1 \rangle$ and $\langle
S_2, I_2, U_2 , \rightarrow_2, \bar{s}_2 \rangle$ be two \iolts{}s with
disjoint sets of input labels $I_1$ and $I_2$, and disjoint sets of
output labels $U_1$ and $U_2$.  The parallel composition of $\bar{s}_1$
and $\bar{s}_2$, denoted $\pcomposition{\bar{s}_1}{\bar{s}_2}$ is the
\iolts $\langle Q, I, U, \rightarrow, \pcomposition{\bar{s}_1}{\bar{s}_2}
\rangle$, where:
\begin{itemize}
\item $Q = \{ \pcomposition{s_1}{s_2} ~|~ s_1 \in S_1, s_2 \in S_2 \}$.
\item $I = (I_1 \cup I_2)\setminus (U_1 \cup U_2)$ and $U = U_1 \cup U_2$.
\item $\to \subseteq Q \times (L \cup \{\tau\})  \times Q$ is the
least relation satisfying:
\end{itemize}
\[
\begin{array}{c}
\begin{array}{ccc}
\infer{\pcomposition{s_1}{s_2} \trans{~x~} \pcomposition{s_1'}{s_2}}
      {s_1 \trans{~x~}_1 s_1' \qquad  x \not\in L_2}
&
\qquad
&
\infer{\pcomposition{s_1}{s_2} \trans{~x~} \pcomposition{s_1}{s_2'}}
      {s_2 \trans{~x~}_2 s_2' \qquad  x \not\in L_1}
\end{array}\\ \\
\infer{\pcomposition{s_1}{s_2} \trans{~x~} \pcomposition{s_1'}{s_2'}}
      {s_1 \trans{~x~}_1 s_1' \qquad  s_2\trans{~x~}_2 s_2' \qquad x\not=\tau}
\end{array}
\]

\end{definition}
The interaction \emph{between} components is
typically intended to be unobservable by a tester. This is not enforced by
the parallel composition, but can be specified by combining parallel
composition with a \emph{hiding} operator, which is formalized below.

\begin{definition}[hiding]\label{def:hiding}
Let $\langle S, I, U, \rightarrow, \bar{s} \rangle$ be an \iolts, and let
$V \subseteq U$. The \iolts resulting from hiding events from the set $V$,
denoted by $\hide{V}{s}$ is the \iolts $\langle S, I, U\setminus V, \to',
\bar{s} \rangle$, where $\to'$ is defined as the least relation satisfying:
\[
\begin{array}{ccccc}
\infer{\hide{V}{s} \trans{~x~}' \hide{V}{s'}}
      {s \trans{~x~} s' \qquad  x \not\in V}
&
\qquad
&
\infer{\hide{V}{s} \trans{~\tau~}' \hide{V}{s'}}
      {s \trans{~x~} s' \qquad x \in V}
\end{array}
\]
\end{definition}
Note that the hiding operator may turn non-divergent \iolts{}s into
divergent \iolts{}s.  As divergence is excluded from the \ioco testing
theory, we must assume such divergences are not induced by composing two
implementations in parallel and hiding all successful communications.
Since implementations are assumed to be input enabled, this can only
be ensured whenever components that are put in parallel never produce
infinite, uninterrupted runs of outputs over their alphabet of shared
output actions. Implementations adhering to these constraints are referred
to as \emph{shared output bounded} implementations. From hereon, we
assume that all the implementions considered are shared output bounded.

\section{Decomposibility}\label{sec::decompos}

Software can be constructed by decomposing a specification of the software
in specifications of smaller complexity. Reuse of readily
available and well-understood platforms or environments can steer such
a decomposition. Given the prevalence of such platforms, the software
engineering and associated testing problem thus shifts to finding a
proper specification of the system from which the platform behavior has
been factored out. Whether this is possible, however, depends on the
specification; if so, we say that a specification is \emph{decomposable}.

The decomposability problem requires known action alphabets for both the
specification and the platform.
Hence, we first fix these alphabets and illustrate how these
are related.  Hereafter, $L_s$ denotes the action alphabet of the
specification $\bar{s}$ and $L_e$ denotes the action alphabet of the
platform $\bar{e}$. The actions of $L_e$ not exposed to $\bar{s}$ are
contained in action alphabet $L_v$, \ie, we have $L_v = L_e \setminus
L_s$. The action alphabet of the quotient will be denoted by $L$, \ie $L =
(L_s\setminus L_e)\cup L_v$.  The relation between the above alphabets
is illustrated in Figure \ref{fig:structure} in the introduction.



\begin{definition}[Decomposability]\label{def:decompoability}
Let $\bar{s} \in \IOLTS{I_s, U_s}$ be a specification, and let $\bar{e}
\in \IOTS{I_e, U_e}$ be an implementation. Let $L_v = I_v \cup U_v$ be a
set of actions of $\bar{e}$ not part of $\bar{s}$. 
Specification $\bar{s}$ is said to be
\emph{decomposable} for \iots $\bar{e}$ iff there is some
specification $\bar{s}' \in
\IOLTS{(I_s\setminus I_e) \cup I_v, (U_s\setminus U_e) \cup U_v}$ for which
both:
\begin{itemize}
\item $\exists \bar{c}\in \IOTS{(I_s\setminus I_e) \cup I_v, (U_s\setminus U_e) \cup U_v}~\st
  \bar{c} \IOCO{} \bar{s}'$, and
\item $
   \forall \bar{c}\in \IOTS{(I_e\setminus I_e) \cup I_v, (U_e\setminus U_e) \cup U_v}~\st
  \bar{c} \IOCO{} \bar{s}'
  \Longrightarrow \hide{L_v}{\pcomposition{\bar{c}}{\bar{e}}} \IOCO{} \bar{s}$
\end{itemize}
\end{definition}
Decomposability of a specification $\bar{s}$ essentially ensures that a
specification $\bar{s}'$ for a subcomponent exists that guarantees that every
\ioco-correct implementation of it is also guaranteed to work correctly
in combination with the platform.
\begin{figure}[ht]
\centering
\subfigure[\iolts $\bar{s}$]{
\label{fig:FIG_s}
\begin{tikzpicture}[->,>=latex,shorten >=1pt,auto,node distance=1 cm,thick, inner sep=1pt, font=\scriptsize]
    \tikzstyle{every state}=[minimum size=0.2mm]
          \node[state]         (0)                    {$\bar{s}$};
          \node[state]         (1) [below right of=0]       {};
          \node[state]         (3) [below   of=1]        {};
          \node[state]          (4)[below left  of=0]        {$s_1$};
          \node[state]          (7)[below of = 4]   {};
          \node[draw=none]   (blank0) [right of=4,xshift=1cm] {};
          \path
                (0) edge                node {c}   (1)
                (1) edge                node {r}   (3)
                    edge                    node[above]{$\tau$}    (4)
                (4) edge                    node{t}    (7)
                    edge                    node{$\tau$}            (0)
                    ;
    \end{tikzpicture}
}
\subfigure[SA $\Delta(\bar{s})$]{
\label{fig:FIG_delta_s}
\begin{tikzpicture}[->,>=latex,shorten >=1pt,auto,node distance=1 cm,thick, inner sep=1pt, font=\scriptsize]
    \tikzstyle{every state}=[minimum size=0.2mm]
          \node[state]         (0)                    {$\bar{q}$};
          \node[state]         (1) [right of=0]       {};
          \node[state]         (3) [below   of=1]        {};
          \node[state]          (2)[right    of=1]        {};
          \node[draw=none]   (blank0) [right of=4,xshift=1cm] {};
          \path
                (0) edge[bend right]                 node {c}   (1)
                    edge[loop left]                     node{$\delta$}        (1)
                (1) edge                node {r}   (3)
                    edge[bend right]     node[above]{$\delta$}    (0)
                    edge                node{t}    (2)
                (2) edge[loop below]          node {$\delta$}        (2)
                (3) edge[loop left]          node {$\delta$}        (3)
                    ;
    \end{tikzpicture}
}
\subfigure[IOTS $\bar{e}$]{
\label{fig:FIG_e}
\begin{tikzpicture}[->,>=latex,shorten >=1pt,auto,node distance=1 cm,thick, inner sep=1pt, font=\scriptsize, bend angle=55]
    \tikzstyle{every state}=[minimum size=0.2mm]
          \node[state]         (0)                          {$\bar{e}$};
          \node[state]         (1) [right   of=0]       {};
          \node[state]         (2) [right   of=1]       {};
          \path
                (0) edge[loop left]         node {error}       (0)
                    edge                    node{c}            (1)
                (1) edge                    node {order}     (2)
                    edge[bend left]         node{$\tau$}      (0)
                (2) edge[bend right ]       node{$\tau$}          (0);
                ;
    \end{tikzpicture}
}
\subfigure[IOTS $\bar{r}$]{
\label{fig:FIG_r}
\begin{tikzpicture}[->,>=latex,shorten >=1pt,auto,node distance=1 cm,thick, inner sep=1pt, font=\scriptsize, bend angle=55]
    \tikzstyle{every state}=[minimum size=0.2mm]
          \node[state]         (0)                          {$\bar{r}$};
          \node[state]         (1) [right   of=0]       {};
          \node[state]         (2) [right   of=1]       {};
          \node[state]         (3) [below  of = 2]       {};
          \path
                (0) edge[loop left]         node {error}       (0)
                    edge                    node{c}            (1)
                (1) edge                    node {order}     (2)
                    edge[bend left]         node{$\tau$}      (0)
                (2) edge[bend right ]       node{$\tau$}          (0)
                    edge[bend right]                 node[left]{error}     (3)
                (3) edge[bend left]                    node{r}         (0)
                    edge[bend right]       node[right]{$\tau$}          (2)
                ;
    \end{tikzpicture}
}
\\
\subfigure[IOLTS $\bar{m}$]{
\label{fig:FIG_m}
\begin{tikzpicture}[->,>=latex,shorten >=1pt,auto,node distance=1 cm,thick, inner sep=1pt, font=\scriptsize]
    \tikzstyle{every state}=[minimum size=0.2mm]
          \node[state]         (0)                    {$\bar{m}$};
          \node[state]         (1) [below  right of=0]       {};
          \node[state]         (3) [below left  of=1]        {};
          \node[state]          (4)[below  right    of=1]        {};
          \node[draw=none]   (blank0) [right of=1,xshift=1cm] {};
          \path
                (0) edge                      node{order}         (1)
                (1) edge                      node {error}    (3)
                    edge                      node {t}      (4)
                    ;
    \end{tikzpicture}
}
\subfigure[IOLTS $\bar{p}$]{
\label{fig:FIG_p}
\begin{tikzpicture}[->,>=latex,shorten >=1pt,auto,node distance=1 cm,thick, inner sep=1pt, font=\scriptsize]
    \tikzstyle{every state}=[minimum size=0.2mm]
          \node[state]         (0)                    {$\bar{p}$};
          \node[state]         (1) [below right of=0]       {};
          \node[state]         (3) [below left  of=1]        {};
          \node[state]          (4)[below  right   of=1]        {};
          \path
                (0) edge                      node {order}         (1)
                (1) edge                      node {error}    (3)
                     edge                      node {t}      (4)
                (3) edge[loop left]                      node {error}    (3)
                    edge                       node {$\tau$}   (0)
                    ;
    \end{tikzpicture}
}
\subfigure[IOTS $\bar{c}$]{
\label{fig:FIG_c}
\begin{tikzpicture}[->,>=latex,shorten >=1pt,auto,node distance=1 cm,thick, inner sep=1pt, font=\scriptsize]
    \tikzstyle{every state}=[minimum size=0.2mm]
          \node[state]         (0)                    {$\bar{c}$};
          \node[state]         (1) [below right of=0]       {};
          \node[state]         (3) [below left   of=1]        {};
          \node[state]          (4)[below right  of=1]        {};
          \path
                (0) edge                      node{order}       (1)
                (1) edge                      node {error}           (3)
                    edge                      node {t}                   (4)
                    edge[loop right]                node{order}                       (0)
                (3) edge[loop  left]                node{order}                       (3)
                    edge                             node{error}                  (0)
                (4) edge[loop right]                node{order}                       (4)
                ;
    \end{tikzpicture}
}
\caption{A specification of a vending machine ($\bar{s}$),
 two behavioral models of an implemented money component ($\bar{e}$ and
 $\bar{r}$) and
 two specifications for a drink component ($\bar{m}$ and
 $\bar{p}$) with the behavioral model of
 an implementation of the drink component ($\bar{c}$).}
\label{fig:FIG}
\end{figure}
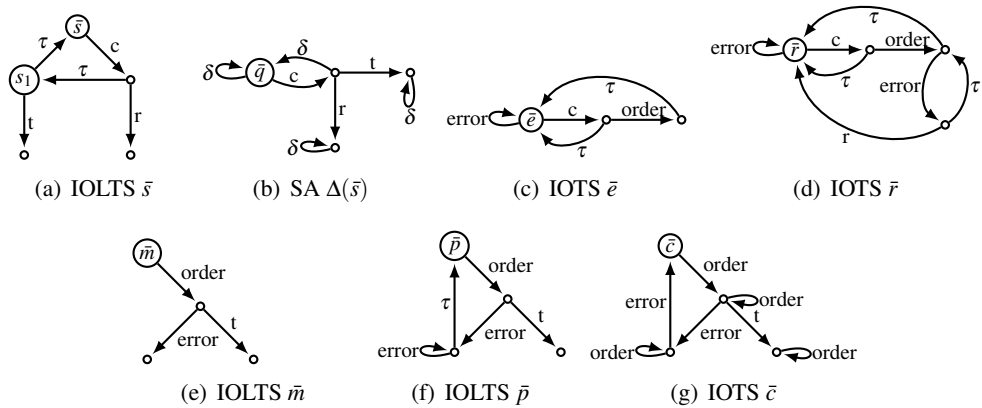
\begin{example}
 Consider \iolts{}s depicted in Figure \ref{fig:FIG}.  The \iots
 $\bar{e}$ \ref{fig:FIG_e} presents the behavioral model of an environment which after
 receiving a coin (c) either orders drink (order) or does nothing.
 Upon receiving an error signal (error), never refunds 
 the money (r). Component $\bar{e}$ interacts with another component
 through actions `order' and `error'; together, the components implement
 a vending machine for which \iolts $\bar{s}$ \ref{fig:FIG_s} is the specification.
 The \iolts $\bar{m}$ \ref{fig:FIG_m} is a specification of a drink component which
 delivers tea after receiving a drink order.  If it encounters a
 problem in delivering the drink, it signals an error.  Specification
 $\bar{m}$ guarantees that the combination of component $\bar{e}$
 with any drink component implementation conforming to $\bar{m}$, also
 conforms to $\bar{s}$.
\end{example}

It may, however, be the case that
an implementation, in combination with a given platform,
perfectly adheres to the overall specification $\bar{s}$, and, yet fails to
pass the conformance test for $\bar{s}'$. As a consequence, non-conformance
of an implementation to $\bar{s}'$ may not by itself be a reason to
reject the implementation.

\begin{example}
Consider \iolts{}s in Figure \ref{fig:FIG}. The \iolts $\bar{m}$ \ref{fig:FIG_m} is a witness for decomposability of \iolts $\bar{s}$\ref{fig:FIG_s} for platform $\bar{e}$\ref{fig:FIG_e}. Thus, any compound system built of \iots $\bar{e}$ and a component conforming to $\bar{m}$  is guaranteed to be in conformance with \iolts $\bar{s}$. Now, consider \iots $\bar{c}$ \ref{fig:FIG_c} which incorrectly
 implements the functionality specified in \iolts $\bar{m}$ \ref{fig:FIG_m}, as it sends
 `error' twice.  Observe that, nevertheless, $\hide{\{ \text{error},
 \text{order}\}}{\pcomposition{\bar{c}}{\bar{e}}}$ still
 conforms to $\bar{s}$.
\end{example}

It is often desirable to consider specifications $\bar{s}'$ for which
one \emph{only} has to check whether an implementation $\bar{c}$
adheres to $\bar{s}'$, \ie, specifications for which it is guaranteed
that a failure of an implementation $\bar{c}$ to comply to $\bar{s}'$
also guarantees that the combination $\pcomposition{\bar{c}}{\bar{e}}$
will violate the original specification $\bar{s}$.  We can obtain this
by considering a stronger notion of decomposability.
\begin{definition}[Strong Decomposability]\label{def:strong_decompoability}
Let $\bar{s} \in \IOLTS{I, U}$ be a specification, and let $\bar{e}
\in \IOTS{I_e, U_e}$ be an implementation. Let $L_v = I_v \cup U_v$ be a
set of actions of $\bar{e}$ not part of $\bar{s}$. 
Specification $\bar{s}$ is said to be \emph{strongly
decomposable}
for \iots $\bar{e}$ iff there is some specification $\bar{s}' \in
\IOLTS{(I_s\setminus I_e) \cup I_v, (U_s\setminus U_e) \cup U_v}$ for which
both:
\begin{itemize}
\item $\exists \bar{c}\in \IOTS{(I_s\setminus I_e) \cup I_v, (U_s\setminus U_e) \cup U_v}~\st
  \bar{c} \IOCO{} \bar{s}'$, and
\item $
   \forall \bar{c}\in \IOTS{(I_s\setminus I_e) \cup I_v, (U_s\setminus U_e) \cup U_v}~\st
  \bar{c} \IOCO{} \bar{s}'
  \Longleftrightarrow
   \hide{L_v}{\pcomposition{\bar{c}}{\bar{e}}} \IOCO{} \bar{s}$
\end{itemize}
\end{definition}
%
\begin{example}
Consider the \iolts{}s $\bar{p}$ and $\bar{e}$ in Figure \ref{fig:FIG};
specification $\bar{p}$ is such that the combination of component
$\bar{e}$ with any \emph{shared output bounded} component that does not conform to $\bar{p}$, fails
to comply to $\bar{s}$.

\end{example}

\section{Sufficient Conditions for Decomposibility\label{sec::sync}}

Checking whether a given specification is decomposable is a difficult
problem. However, knowing that a specification is decomposable in
itself hardly helps a design engineer. Apart from the question whether a
specification is decomposable, one is typically interested
in a witness for the decomposed specification, or \emph{quotient}. Our approach
to the decomposability problem is therefore constructive: we define a
quotient and we identify several conditions that ensure that the quotient
we define is a witness for the decomposability of a given specification.

One of the problems that may prevent a specification from being
decomposable for a given platform $\bar{e}$ is that the latter may exhibit
some behavior which unavoidably violates the specification $\bar{s}$.
We shall therefore only consider platforms for which such violations
are not present. We formalize this by checking whether the behavior of
$\bar{e}$ is \emph{included} in the behavior of $\bar{s}$; that is, we
give conditions that ensure that $\bar{e}$ in itself cannot violate the
given specification $\bar{s}$. Moreover, we assume that the input-enabled
specification of $\bar{e}$ is available.

Assuming that the behavior of $\bar{e}$ is included in the behavior of
the given specification $\bar{s}$, we then propose a quotient $\bar{s}'$
of $\bar{s}$ for $\bar{e}$ and prove sufficient conditions that guarantee
that $\bar{s}$ is indeed decomposable and $\bar{s}'$ is a witness to that.

\subsection{Inclusion relation}
We say that the behavior of a given platform $\bar{e}$ is included in
a specification $\bar{s}$ if the outputs allowed by $\bar{s}$ subsume
all outputs that can be produced by $\bar{e}$. For this, we need to
take possible communications between $\bar{e}$ and the to-be-derived
quotient over the action alphabet $L_v$ into account. Another issue is that
we are dealing with two components, each of which may be quiescent. If
component $\bar{e}$ is quiescent, its quiescence may be masked by
outputs from the component with which it is supposed to interact. We must
therefore consider a refined notion of quiescence.  We say state $s$
in specification $\bar{s}$ is \emph{relatively quiescent} with respect
to alphabet $L_e$, denoted by $\delta_{\bar{e}}(s)$, if $s$ produces
no output of $L_e$, \ie $\out{s}\cap L_e = \emptyset$. Analogous to
$\delta$, the suspension traces of $\bar{s}$ can be enriched by adding
the rule $s\Trans{\delta_{\bar{e}}}_\delta s$ for $\delta_{\bar{e}}(s)$
to be able to formally reason about the possibility of being relatively
quiescent with respect to $L_e$.  We write $\EStraces{\bar{s}}{\bar{e}}$
to denote this enriched set of suspension traces of $\bar{s}$.

Since the suspension traces of $\bar{s}$ and $\bar{e}$ differ as a
result of different alphabets, we introduce a \emph{projection operator}
which allows us to map the suspension traces of $\bar{s}$ to suspension
traces of $\bar{e}$.  The operator $\projection{\_}{L_e}$ is defined as
$\projection{(x\sigma)}{L_e} = x\projection{\sigma}{L_e}$ if $x\in L_e$; 
$\projection{(x\sigma)}{L_e} = \delta(\projection{\sigma}{L_e})$, if  $x\in\{\delta, \delta_{\bar{e}}\}$;
otherwise, $\projection{(x\sigma)}{L_e} = \projection{\sigma}{L_e}$.

\begin{definition}\label{def:PIOCO}
Let \iots $\tuple{ S_e, I_e, U_e, \rightarrow, \bar{e}}$ be an
implementation. Let \iolts $\tuple{ S_s, I_s, U_s, \rightarrow,
\bar{s}}$ be a specification. 
We say the behavior of $\bar{e}$ is included in $\bar{s}$,
denoted by $\bar{e} \INCL \bar{s}$ iff
\[\forall \sigma\in \EStraces{\bar{s}}{\bar{e}}:
\out{\hide{L_v}{\bar{e}} \after \projection{\sigma}{L_e}} \subseteq \out{\bar{s}
\after \sigma}
\]
\end{definition}
%
\begin{example}
Consider the \iolts{}s in Figure \ref{fig:FIG}. We have $\bar{e}\INCL
\bar{s}$.  Consider the \iolts $\bar{r}$ which has the same functionality with
\iolts $\bar{e}$ except that upon receiving an error signal (error),
it may or may not refund the money (r). The behavior of $\bar{r}$ is not included in $\bar{s}$, because of
observing the output $r$ in $\bar{r}$ after executing $\projection{(ct)}{L_e}$ while $\bar{s}$ after execution of $ct$ reaches to a quiescent state.
\end{example}
\subsection{Quotienting}
We next focus on deriving a quotient of the specification $\bar{s}$,
factoring out the behavior of the platform $\bar{e}$.  A major source of
complexity in defining such a quotient is the possible non-determinism
that may be present in $\bar{s}$ and $\bar{e}$. We largely avert this
complexity by utilizing the suspension automata underlying $\bar{s}$
and $\bar{e}$.


Another source of complexity is the fact that we must reason about the
states of two systems running in parallel; such a system synchronizes
on shared actions and interleaves on non-shared actions. We tame this 
conceptual complexity by formalizing an $\execute$ operator which, 
when executing a shared or non-shared action, keeps track of the set 
of reachable states for the (suspension automata) of $\bar{s}$ and $\bar{e}$. 
Formally, the $\execute$ operator is defined as follows.
\begin{definition} \label{def:execute}
Let $\tuple{Q_s,I_s, U_s\cup \{ \delta\}, \to_s, \bar{q}_s}$ be a
suspension automaton underlying specification \iolts $\bar{s}$, and let
$\tuple{Q_e,I_e, U\cup\{\delta\}, \to_e, \bar{q}_e}$ be a suspension automaton
underlying platform \iolts $\bar{e}$. Let $q \in \powerSet{Q_s \times
Q_e}$ be a non-empty collection of sets and let $x \in L_s\setminus
(L_e\setminus L_v)$.

\[
q \execute x
=  \left\{
      \begin{array}{l l}
      \displaystyle
        \bigcup_{\sigma\in L_e^*} \bigcup_{(s,e)\in q}
        \{(q'_s,q'_e) \mid s \trans{\ \sigma\ }_s q'_s \text{ and } e \trans{\,\sigma x\,}_e q'_e \}
        &  \text{if } x\in L_v\\
      \displaystyle
        \bigcup_{\sigma\in L_e^*}\bigcup_{(s,e)\in q}\{(q_s',q_e') \mid s \trans{\,\sigma x\,}_s q_s' \text{ and } e \trans{\ \sigma\ }_e q_e' \}
         &  \text{if } x\not\in L_v\\
      \displaystyle
        \bigcup_{\sigma\in L_e^*}\bigcup_{(s,e)\in q}\{(q_s',q_e') \mid s \trans{\,\sigma\delta}_s q_s' \text{ and } e \trans{\,\sigma \delta}_e q_e' \}
         &  \text{if } x = \delta\\
  \end{array} \right.
\]
\end{definition}
Using the $\execute$ operator, we have an elegant construction of
an automaton, called a \emph{quotient automaton}, see below, which allows
us to define sufficient conditions for establishing the decomposability
of a given specification.

\newcommand{\inv}[1]{\ensuremath{#1}^{-1}}
\begin{definition}[Quotient Automaton] \label{def:SpecQuotient}
Let $\tuple{ Q_s, I_s, U_s\cup\{\delta\}, \rightarrow_s, \bar{q}_s }
$ be a suspension automaton underlying specification $\bar{s}$, and
let $\tuple{ Q_e, I_e, U_e\cup\{\delta\}, \rightarrow_e, \bar{q}_e}$
be a suspension automaton underlying platform $\bar{e}$. The
quotient of $\bar{s}$ by $\bar{e}$, denoted by $\quotient{\bar{s}}{\bar{e}}$
is a suspension automaton
$\tuple{ Q, I, U\cup\{\delta\}, \rightarrow, \bar{q}}$ where:
\begin{itemize}
\item $Q = (\powerSet{Q_s \times Q_e}\setminus \{\emptyset\}) \cup
Q_\delta$, where $Q_\delta = \{q_\delta\mid q \in \powerSet{Q_s \times
Q_e}, q \not= \emptyset\}$; for $q \notin Q_\delta$, we set $\inv{q}
= q$ and for $q_\delta \in Q_\delta$, we set $\inv{q_\delta} = q$.

\item $\bar{q} = \{(\bar{q}_s,\bar{q}_e)\}$.

\item $I = (I_s\setminus I_e)\cup (U_e\setminus U_s) $ and $U =
(U_s\setminus U_e)\cup \{\delta\}\cup (I_e\setminus I_s)$.

\item $\to \subseteq Q \times L \times Q$ is the least set satisfying:
\end{itemize}
\[
\begin{array}{c}
\\
\begin{array}{ccc}
\infer[\lbrack I_1\rbrack]{q \trans{a} \inv{q} \execute a}
      {a \in I \quad \inv{q} \execute a \not= \emptyset}
&
\qquad
& 
\infer[\lbrack U_1\rbrack]{q \trans{x} \inv{q} \execute x}
      {x \in U_v \quad q \notin Q_\delta \quad
       \inv{q} \execute x \not= \emptyset}
\end{array}
\\
\\
\infer[\lbrack U_2 \rbrack]{q \trans{x} \inv{q} \execute x}
      {x \in U \setminus U_v \quad
       \forall (s,e) \in q, \sigma \in \traces{s} \cap \traces{e} \cap (L_\delta^*\setminus L_\delta^*\delta):~ x \in
       \out{s \after \sigma}
      }
\\
\\
\infer[ \lbrack \delta_1\rbrack ]{q \trans{\delta} \inv{q} \execute \delta}
      {\forall (s,e) \in \inv{q}, \sigma \in \traces{s} \cap \traces{e}:~ \delta \in
       \out{s \after \sigma}
      }
\\
\end{array}
\]
\end{definition}
We briefly explain the construction of a quotient automaton.
A \emph{non-shared input action} is added to a state in the quotient automaton
$\quotient{\bar{s}}{\bar{e}}$ if an execution of the corresponding state
in $\bar{e}$ leads to a state in $\bar{s}$ at which that action is enabled ($I_1$, in combination with the second case in Definition \ref{def:execute}).
A \emph{shared input action} obeys the same rule except that a state of $\bar{e}$
has to be reachable where that input action is taken ($I_1$, in combination with the first case in Definition \ref{def:execute}).
Note that a shared input action of $\quotient{\bar{s}}{\bar{e}}$ is an output action
from the viewpoint of $\bar{e}$.
In contrast, a \emph{non-shared output action} is allowed at a state of
$\quotient{\bar{s}}{\bar{e}}$ only if it is allowed by $\bar{s}$
after any possible execution of $\bar{e}$ ($U_2$) and
a similar rule is applied to quiescence ($\delta_1$).
Analogous to the shared input actions, a \emph{shared output action} is considered as an action
of a state whenever a valid execution of the correspondent states in $\bar{e}$ leads to a state at
which that output action is enabled ($U_1$).
Because the shared actions are hidden in $\bar{s}$, a shared output
action, in $\quotient{\bar{s}}{\bar{e}}$, may also be enabled at a state reached by $\delta$ transitions.
Such a sequence of events is invalid due to the definition of quiescence. 
The observed problem is solved by adding a special set of states $Q_\delta$
to the states of the quotient automaton.
These states represent quiescent states corresponding
to the reachable states after executing $\delta$ in $\quotient{\bar{s}}{\bar{e}}$.
Moreover, no shared output action is added to these states. 
%
\begin{figure}[ht]
\centering
\subfigure[SA $\bar{r}$]{
\begin{tikzpicture}[->,>=latex,shorten >=1pt,auto,node distance=1 cm,thick, inner sep=1pt, font=\scriptsize]
    \tikzstyle{every state}=[minimum size=0.2mm]
          \node[state]         (0)        {$\bar{r}$};
          \node[state]         (1) [below  right of=0]       {};
          \node[state]         (2) [below  left of=0]       {};
          \node[state]         (3) [right  of=1]      {};
          \node[state]         (4) [below  right   of=3]      {};
          \node[state]         (5) [below     of=1]       {1};
          \node[state]         (6) [above right  of= 4]       {};
          \node[state]         (7) [above     of=0]       {1};
          \path
          (0) edge                      node[near start] {order}                    (1)
              edge                      node {error}                    (7)
              edge                      node [near start, left] {$\delta$}                     (2)
          (1) edge                      node {${t}$}             (3)
              edge [bend left]                      node {error}                    (5)
          (2) edge                      node {order}           (1)
              edge [loop left]          node {${\delta}$}          (2)
          (3) edge                      node {${\delta}$}          (4)
              edge                      node {error}                    (6)
          (4) edge [loop left]          node {${\delta}$}          (4)
          (5) edge [bend left]          node[right]{$\delta$}             (2)
              edge [bend left]          node {order}           (1)
              edge [loop right]          node {error}          (5)
          (6) edge [loop right]         node {error}           (6)
              edge                      node {${\delta}$}          (4)
          (7) edge [bend right=50]         node[left] {$\delta$}          (2)
              edge [bend left=50]         node {order}          (1)
              edge [loop right]         node {error}           (7)
          ;
    \end{tikzpicture}
}
\subfigure[SA $\bar{i}$]{
\begin{tikzpicture}[->,>=latex,shorten >=1pt,auto,node distance=1 cm,thick, inner sep=1pt, font=\scriptsize]
    \tikzstyle{every state}=[minimum size=0.2mm]
          \node[state]         (0)                          {$\bar{i}$};
          \node[state]         (1) [below  right of=0]       {};
          \node[state]         (8) [right       of=1]      {};
          \node[state]         (3) [right       of=8]      {};
          \node[state]         (4) [below right of=3]      {};
          \node[state]         (5) [below left  of=1]       {};
          \node[state]         (6) [above right of=4]       {};
          \node[state]         (7) [below left  of=0]       {};
          \node[state]         (9) [below   of=8]       {};
          \path
          (0) edge                      node {order}                    (1)
              edge                     node[left] {error}                    (7)
          (1) edge                      node {t}             (8)
              edge                      node {error}                    (5)
          (3) edge                      node {$\delta$}          (4)
              edge                      node {error}                    (6)
          (4) edge [loop left]          node {$\delta$}          (4)
          (5) 
              edge                      node {error}          (7)
          (6) edge [loop right]         node {error}           (6)
              edge                      node {$\delta$}          (4)
          (7) edge [loop left]          node {error}          (7)
              edge                      node {order}                    (1)
          (8) edge                      node {error}          (3)
              edge                      node {$\delta$}          (9)
         (9) edge [loop left]           node {$\delta$}          (9)
          ;
    \end{tikzpicture}
}
\caption{Two quotient automata derived using Definition \ref{def:SpecQuotient}}
\label{fig:QA}
\end{figure}
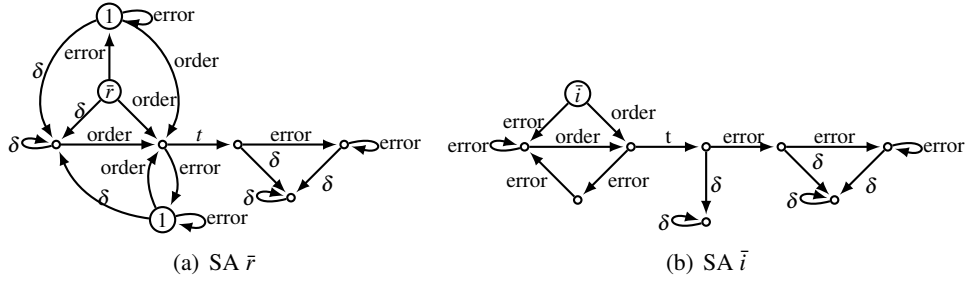

%
The quotient automaton derived from specification $\bar{s}$ and platform
$\bar{e}$ is a suspension automaton: it is deterministic and it has
explicit $\delta$ labels. Yet, the quotient automata we derive are not
necessarily valid suspension automata.
(As we recalled in Section~\ref{sec::pre},
only \emph{valid} suspension automata have the same testing power
as ordinary \iolts{}s.)  We furthermore observe that there some
quotient automata that are valid suspension automata but nevertheless
only admit non-shared output bounded implementations as implementations
that conform to the quotient. As observed earlier, such implementations 
unavoidably give rise to divergent systems when composed in parallel with
the platform.
\begin{example}
  Consider SA{}s depicted in Figure \ref{fig:QA}, \iolts{}s $\bar{s}$ and $\bar{e}$
  in Figure \ref{fig:FIG} and \iolts $\bar{l}$ derived by removing
  the internal transition from state $s_1$ to the initial state in $\bar{s}$.
  SA $\bar{r}$ is the quotient of $\bar{s}$ by $\bar{e}$. Likewise, SA $\bar{i}$ is the
  quotient of $\bar{l}$ by $\bar{e}$.
 Suspension automata $\bar{r}$ and $\bar{i}$ are valid SA regarding
  the definition of validity of suspension automata presented in \cite{Willemse} .
  Assume an arbitrary shared output bounded \iots $\bar{c}$ whose length of the longest
  sequence on the shared output is $n$
 , \ie $\out{\bar{c}\after \sigma}\subseteq \{tea,\delta\} \text{ for } \sigma= \{error\}^n$.
  Clearly, $\bar{c} ~\NIOCO{}~ \bar{i}$, because $\out{\bar{i}\after \sigma}=\{error\}$.
  However, for any $n\ge 0$, there is always a shared output bounded \iots that conforms to $\bar{r}$.
\end{example}
%
%
In view of the above, we say that a quotient automaton is \emph{valid}
if it is a valid suspension automaton and strongly non-blocking.
\begin{definition}
Let $\quotient{\bar{s}}{\bar{e}}$ be a quotient automaton derived from
a specification $\bar{s}$ and an environment $\bar{e}$. We say that
$\quotient{\bar{s}}{\bar{e}}$ is \emph{valid} iff both:
\begin{itemize}
\item $\quotient{\bar{s}}{\bar{e}}$ is a valid suspension automaton, and
\item $\quotient{\bar{s}}{\bar{e}}$ is strongly non-blocking, \ie
$\forall q\in \quotient{\bar{s}}{\bar{e}}\st \out{q}\cap ((U\setminus U_v)\cup\{\delta\})\neq \emptyset$.
\end{itemize}
\end{definition}
%
%
Strongly non-blocking ensures
that the quotient automaton always admits a shared output bounded implementation that conforms to it.
Furthermore, valid quotient automata are, by definition,
also valid suspension automata. Since every valid suspension automaton
underlies at least one \iolts, we therefore have established a sufficient
condition for the decomposability of a specification. %
\begin{theorem}\label{theo:decomposability}
Let $\bar{s} \in \IOLTS{I_s, U_s}$ be a specification and let $\bar{e}
\in \IOTS{I_e, U_e}$ be an environment. Then $\bar{s}$ is decomposable
for $\bar{e}$ if $\quotient{\bar{s}}{\bar{e}}$ is a valid quotient
automaton and $\bar{e} \INCL \bar{s}$.
\end{theorem}
Note that the \iolts underlying the quotient automaton is a witness
to the decomposability of the specification; we thus not only have a
sufficient condition for the decomposability of a specification but also
a witness for the decomposition.


\subsection{Example}\label{sec:example}
\begin{figure}[t]
\centering
\subfigure[Schematic view]{
\includegraphics[width=5cm]{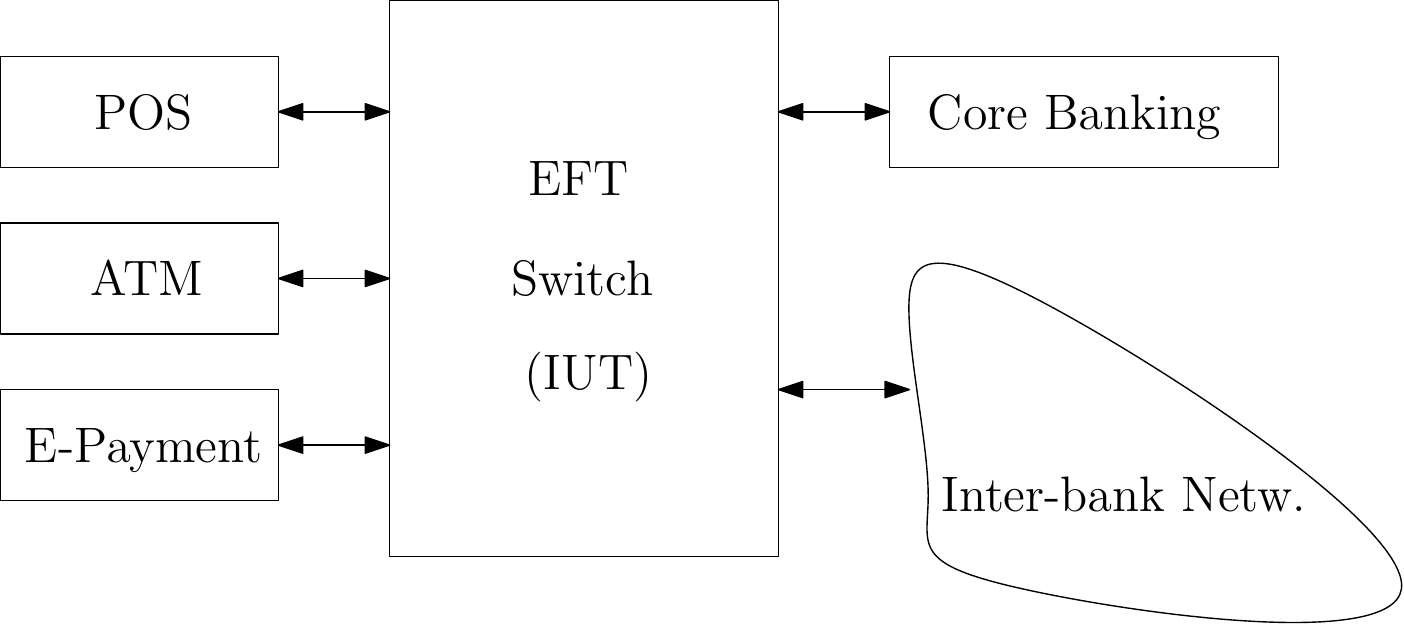} 
\label{fig::EFTSchematic_eft}
}
\subfigure[SA $\bar{s}$]{
 \begin{tikzpicture}[->,>=latex,shorten >=1pt,auto,thick, inner sep=1pt, font=\scriptsize, shape=circle]
          \node[draw]         (0)                       {$\bar{s}$};
          \node[draw]         (1) [above right of=0]    {1};
          \node[]             (2) [right of=1]    {};
          \node[draw]         (3) [right of= 2]         {3};
          \node[draw]         (4) [right of= 3]         {4};
          \node[draw]        (5) [below right of = 0]  {5};
          \node[]        (6) [right of = 5]  {};
          \node[draw]        (7) [right of = 6]  {7};
          \path
                (0) edge              node  {p\_rq}   (1)
                    edge              node[left]  {rev\_rq}   (5)
                (1) edge              node {rev\_rq} (3)
                    edge[bend right]  node[below] {p\_rs} (3)
                (3) edge              node {$\tau$} (4)
                    edge[loop below]      node[right] {rev\_rq} (3)
                (5) edge[loop above]      node[right] {p\_rq} (5)
                    edge[loop below]      node[left] {rev\_rq} (5)
                    edge              node[below] {$\delta$} (7)
                (7) edge[loop right]      node[right] {$\delta$} (7)
                    ;
\end{tikzpicture}
\label{fig::EFTSchematic_s}
}\\
\subfigure[SA $\bar{e}$]{
 \begin{tikzpicture}[->,>=latex,shorten >=1pt,auto,node distance=1 cm,thick, inner sep=1pt, font=\scriptsize, shape=circle]
          \node[draw]         (0)                       {$\bar{e}$};
          \node[draw]         (1) [right of=0]    {1};
          \node[draw]         (2) [above right of=1]    {2};
          \node[draw]         (3) [below right of= 1]   {3};
          \path
                (0) edge              node  {p\_rq}    (1)
                    edge[loop left]   node  {p\_rs}    (0)
                (1) edge              node[left] {p\_rs}     (3)
                    edge              node {t} (2)
                (2) edge[loop right]        node {p\_rs, $\delta$} (2)
                (3) edge[loop right]        node {p\_rs, $\delta$} (3)
                ;
\end{tikzpicture}
\label{fig::EFTSchematic_e}
}
\subfigure[$\quotient{\bar{s}}{\bar{e}}$]{
 \begin{tikzpicture}[->,>=latex,shorten >=1pt,auto,node distance=1.2 cm,thick, inner sep=1pt, font=\scriptsize, shape=circle]
          \node[draw]         (0)                       {0};
          \node[draw]         (1) [above right of=0]    {1};
          \node[draw]         (2) [below right of=0]    {2};
          \node[draw]         (3) [right of= 1]         {3};
          \node[draw]         (4) [right of= 2]         {4};
          \node[draw]         (5)  [right of = 3]  {5};
          \node[draw]         (6)  [right of = 4]  {6};
          \node[draw]         (7)  [right of = 6]  {7};

          \path
                (0) edge              node  {t}   (1)
                    edge              node[left]  {rev\_rq}     (2)
                (1) edge              node  {rev\_rq} (3)
                (2) edge[bend left = 80]   node[right] {t} (6)
                    edge              node[below] {rev\_rq} (4)
                (3) edge              node {$\delta$} (5)
                    edge[loop below]  node[left] {rev\_rq} (3)
                (4) edge              node[below] {t} (6)
                    edge[loop above]  node[left] {rev\_rq} (4)
                (5) edge [loop right] node[right] {$\delta$} (5)
                (6) edge              node {$\delta$} (7)
                    edge[loop above]  node[right] {rev\_rq} (6)
                (7) edge [loop right] node[right] {$\delta$} (7)
          ;
\end{tikzpicture}
\label{fig::EFTSchematic_quotient}
}
\caption{A schematic view of the EFT Switch, a suspension automata of
simplified behavioral models of the EFT switch $\bar{s}$ and an implementation of the financial component $\bar{e}$,
and the quotient of $\bar{s} $ w.r.t.\ $\bar{e}$}
\label{fig::EFTSchematic}
\end{figure}
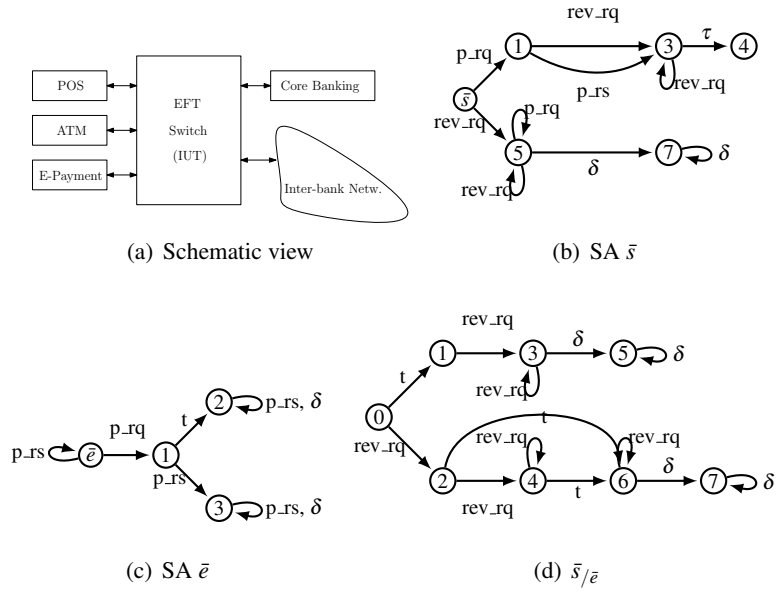

To illustrate the notions introduced so far, we treat a simplified model of an Electronic Funds Transfer (EFT) switch, which we have studied and tested using ioco-based techniques \cite{AsaadiKMN11}. A schematic view of this example is depicted in Figure \ref{fig::EFTSchematic_eft}.
An EFT switch provides a communication mechanism among different components of a card-based financial system.
On one side of the EFT switch, there are components, with which the end-user deals, such as
Automated Teller Machines (ATMs), Point-of-Sale (POS) devices and e-Payment applications.
On the other side, there are banking systems and the inter-bank network connecting
the switches of different financial institutions.

The various involving parties in every transaction performed by an EFT switch
in conjunction with the variety of financial transactions complicate the behavioral model of the EFT switch. Similar to any other complex software system, the EFT switch comprises
many different components, some of which can be run individually.

A part of the simplified communication model of the EFT switch with a
banking system in the purchase scenario is depicted in Figure \ref{fig::EFTSchematic_s}.
The scenario starts by receiving a purchase request from a POS; this initial part of the scenario is removed from the model,
 for the sake of brevity. Subsequently, the EFT switch sends a purchase request ($p\_rq$) to the banking system.
The EFT switch will reverse ($rev\_rq$) the sent purchase request if the corresponding response ($p\_rs$) is
not received within a certain amount of time (e.g, an internal time-out occurs, denoted by $\tau$).
Due to possible delays in the network layer of the EFT switch, an external observer (tester)
may observe the reverse request of a purchase even before the purchase request which is pictured
in Fig \ref{fig::EFTSchematic_s}. 

The EFT switch is further implemented in terms of two components, namely, the financial component and the reversal component.
A simplified behavioral model of the financial component is given in Figure \ref{fig::EFTSchematic_e}.
Comparing the two languages of $\bar{s}$ and
$\bar{e}$, $t$ action (representing time-out) is considered as an internal interface between $\bar{e}$ and
a to-be-developed implementation of the reversal component. Observe
that for every sequence
$\sigma$ in $\{p\_rq(\delta_e|rev\_rq)^*, p\_rq(\delta_e|rev\_rq)^* rev\_rq (\delta|\delta_e)^*,$
               $p\_rq ~ p\_rs (\delta_e|rev\_rq)^* (\delta|\delta_e)^*,$
               $(\delta_e|rev\_rq)^*, (\delta_e|rev\_rq)^* rev\_rq(rev\_rq|p\_rq)^* (\delta|\delta_e)^*\}$,
it holds that ~~$\out{\hide{t}{\bar{e}} ~~\after \projection{\sigma}{L_e}}$~~ $\subseteq$ $\out{\bar{s} \after \sigma}$;
thus, the behavior of $\bar{e}$ is included in $\bar{s}$. We next instigate
investigate decomposability of $\bar{s}$ with $\bar{e}$, by constructing
the quotient $\quotient{\bar{s}}{\bar{e}}$. Note
that $t$ is the only shared action which is an input action
from the view point of $\quotient{\bar{s}}{\bar{e}}$. The
resulting quotient automaton, obtained by applying
Definition~\ref{def:SpecQuotient} to $\bar{s}$ and $\bar{e}$ is
depicted in Figure~\ref{fig::EFTSchematic_quotient}. We illustrate some
steps in its derivation.  The initial state of the quotient automaton is
defined as the $\{(\bar{s}, \bar{e})\}$. Below, we illustrate which of the
rules of Definition~\ref{def:SpecQuotient} are possible from this initial
state; doing so repeatedly for all reached states will ultimately produce
the reachable states of the quotient automaton.

\begin{enumerate}
  \item We check the possibility of adding input transitions to the initial state, \ie $q_0 = \{(\bar{s}, \bar{e})\}$.
  Following $q_0 \execute t = \{(s_1, e_2)\}$ and deduction rule $I_1$
  in Definition \ref{def:SpecQuotient}, the transition $q_0\trans{t}q_1$ is
  added to the transition relation of $\quotient{\bar{s}}{\bar{e}}$ where $q_1 = \{(s_1, e_2)\}$ (state  1 in Figure \ref{fig::EFTSchematic_quotient}).
  \item We check the possibility of adding output transitions to $q_0 = \{(\bar{s}, \bar{e})\}$.
  We observe that $rev\_rq\in \out{\bar{s} \after \sigma}$ for every
  $\sigma\in \{\ep, p\_rq, p\_rq~p\_rs\}$. Regarding deduction rule $U_2$, the transition
  $q_0\trans{rev\_rq}q_2$ is added to the transition relation of $\quotient{\bar{s}}{\bar{e}}$
  where $q_2 = \{(s_5, \bar{e}), (s_2, e_1), (s_2, e_3)\}$ (state 2 in Figure \ref{fig::EFTSchematic_quotient}).
  \item Following deduction rule $\delta_1$ and $\delta\not\in \out{\bar{s}\after \ep}$, $\delta$-labeled transition is not added to $q_0$.
\end{enumerate}
The constructed quotient automaton $\quotient{\bar{s}}{\bar{e}}$ is valid:
it is both a valid suspension automaton and strongly non-blocking. As a
result, $\bar{s}$ is decomposable with respect to $\bar{e}$ and
$\quotient{\bar{s}}{\bar{e}}$ is a witness to that.

\section{Strong Decomposibility\label{sec::strong_decompos}}
It is a natural question whether the quotient automaton that
we defined in the previous section, along with the sufficient
conditions for decomposability of a specification provide
sufficient conditions for strong decomposability. The proof of
Theorem~\ref{theo:decomposability} gives some clues to the
contrary. A main problem is in the notion of quiescence, and, in particular
in the notion of \emph{relative} quiescence, which is unobservable in the
standard \ioco theory. More specifically, the platform $\bar{e}$ may
mask the (unwanted) lack of outputs of the quotient automaton.

A natural solution to this is to consider a subclass of implementations
called \emph{internal choice} \iots{}s, studied in~\cite{Neda11,WeiglhoferW09}:
such implementations \emph{only} accept inputs when reaching a quiescent
state. The proposition below states that strong decomposability can be
achieved under these conditions.
\begin{theorem}
\label{theo:decomposability_2}
Let $\bar{s} \in \IOLTS{I_s, U_s}$ be a specification and let
$\bar{e} \in \IOTS{I_e, U_e}$ be an environment. If $\bar{s}$ is
decomposable and $\bar{e}$ is an internal choice \iots then
$\bar{s}$ is \emph{strongly decomposable} and $\quotient{\bar{s}}{\bar{e}}$
is a witness to this.
\end{theorem}
As a result of the above theorem, testing whether the composition of
a component $\bar{c}$ and a platform $\bar{e}$ conforms to
specification $\bar{s}$ reduces to testing for the conformance of
$\bar{c}$ to $\quotient{\bar{s}}{\bar{e}}$. This can be done using the
standard \ioco testing theory~\cite{Tretmans08}.

A problem may arise when trying this approach in practice. Namely,
the amount of time and memory needed for derivation of the \ioco test
suit increases
exponentially in the number of transitions in the specification
due to the nondeterministic nature of the test-case generation algorithm.
We avoid these complexities by presenting an on-the-fly testing algorithm inspired by \cite{VriesT00}.
Algorithm \ref{alg:on-the-fly} describes the on-the-fly testing algorithm in which
sound test cases are generated without constructing the quotient automaton upfront.
We partially explored the quotient automaton during test execution.
We use the extended version of $\execute$ operator in Algorithm \ref{alg:on-the-fly} which is defined
on ordinary \iolts{}s; the underlying \iolts{}s of suspension automata is used to avoid the
complexity of constructing suspension automata, \ie
$\execute: \powerSet{\powerSet{S_s} \times \powerSet{S_e}} \times L_\delta  \times \powerSet{\powerSet{S_s}\times \powerSet{S_e}}$.
\begin{algorithm}
\label{alg:on-the-fly}
Let $\bar{s} \in \IOLTS{I_s, U_s}$ be a specification and let
$\bar{e} \in \IOTS{I_e, U_e}$ be an environment.
Let $\bar{c}\in \IOTS{L_I, L_U}$ be
an implementation tested against $\bar{s}$ with respect to $\bar{e}$
by application of the following rules,
initializing $S$ with $(\{(\bar{s}\after \ep), (\bar{e}\after \ep)\})$ and
verdict $V$ with \emph{None}:\\
while ($V \not\in \{\emph{Fail},\emph{Pass}\}$) \\
\{ apply one of the following case:
\begin{enumerate}
  \item (*provide an input*)~
  Select an $a\in \{a\in L_I\mid S \execute a \neq \emptyset\}$,
  then $S = S \execute a$ and provide $\bar{c}$ with $a$
  \item (*accept quiescence*)~
  If no output is generated by $\bar{c}$ (quiescence situation) and \\
  ${\forall (s,e) \in S, \sigma \in \Straces{s} \cap \Straces{e}:~ \delta \in
       \out{s \after \sigma}}$
      , then $S = S\execute \delta$
  \item (*fail on quiescence*)\label{item:no_quiescence}~
  If no output is generated by $\bar{c}$ (quiescence situation) and\\
  $({\exists (s,e) \in S, \sigma \in \Straces{s} \cap \Straces{e}:~ \delta \not\in
       \out{s \after \sigma}})$, then $V = \emph{Fail}$
  \item (*accept a shared output*)\label{item:no_divergence}~
  If $x\in U_v$ is produced by $\bar{c}$ and $S\execute x\neq \emptyset$,
  then $S = S\execute x$
  \item (*fail on a shared output*)\label{item:no_uv}~
  If $x\in U_v$ is produced by $\bar{c}$ and $S\execute x = \emptyset$,
  then $V = \emph{Fail}$
  \item (*accept an output*)~
  If $x\in U\setminus U_v$ is produced by $\bar{c}$ and
  $\forall (s,e) \in S, \sigma \in \Straces{s} \cap \Straces{e}\cap (L_\delta^*\setminus L_\delta^*\delta):~ x \in
       \out{s \after \sigma}$, then $S = S\execute x$
  \item (*fail on an output*)\label{item:no_u}~
  If $x\in U\setminus U_v$ is produced by $\bar{c}$ and\\
  $\exists (s,e) \in S, \sigma \in \Straces{s} \cap \Straces{e}\cap (L_\delta^*\setminus L_\delta^*\delta):~ x \not\in
       \out{s \after \sigma}$, then $V = \emph{Fail}$
  \item (*nondeterministically terminate*)~
  $V = \emph{Pass}$\quad
\}
\end{enumerate}
\end{algorithm}
Termination of the above algorithm with $V= \emph{Fail}$ implies that
the composition of the implementation under test with $\bar{e}$
does not conform to $\bar{s}$.
\begin{theorem}
\label{theo:decomposability_3}
Let $\bar{s} \in \IOLTS{I_s, U_s}$ be a specification and let
$\bar{e} \in \IOLTS{I_e, U_e}$ be an internal choice \iots environment
whose behavior is included in $\bar{s}$. Let
$V$ be the verdict upon termination
of Algorithm \ref{alg:on-the-fly} when executed on an implementation $\bar{c}$.
If
$\hide{L_v}{\pcomposition{\bar{c}}{\bar{e}}} \IOCO{} \bar{s}$ then
$V = \emph{Pass}$.
%
%
\end{theorem}

\section{Conclusions\label{sec::conc}}
We investigated the property of \emph{decomposability} of a specification
in the setting of Tretmans' \ioco theory for formal conformance
testing~\cite{Tretmans96}.  Decomposability allows for determining
whether a specification can be met by some implementation running on a
given platform.  Based on a new specification, to which we refer to as
the \emph{quotient}, and which we derived from the given one by factoring
out the effects of the platform, we identified three conditions (two
on the quotient and one on the platform) that
together guarantee the decomposability of the original specification.

Any component that correctly implements the quotient is guaranteed to
work correctly on the given platform. However, failing implementations
provide no information on the correctness of the cooperation between
the component and the platform. We therefore studied \emph{strong
decomposability}, which further strengthens the decomposability problem to
ensure that only those components that correctly implement the quotient
are guaranteed to work correctly on the given platform, meeting the
overall specification. This ensures that testing a component against
the quotient provides all information needed to judge whether it will
work correctly on the platform and meet the overall specification's
requirements. However, the complexity of computing the quotient is an
exponential problem. We propose an on-the-fly
test case derivation algorithm which does not compute the quotient
explicitly.  Components that fail such a test case provably fail to
work on the platform, meeting the overall specification, too.

Checking the inclusion relation of a platform may be expensive in practice.
 As for future work,
we would like to merge the two steps of checking the correctness of the platform and driving the quotient and investigate whether the constraints on the platform can be relaxed by ensuring that the derived quotient masks some of the
unwanted behavior of the platform.

\bibliographystyle{eptcs}
\bibliography{bibliography}


\end{document}